\documentclass[final]{svjour2}
\usepackage{epsfig}
\usepackage{amsmath}
\usepackage{amssymb}
\usepackage{amsfonts}
\usepackage{mathptmx}
\usepackage{eucal}
\usepackage{bm}
\usepackage{rotating}
\usepackage[numbers]{natbib}
\makeatletter
\journalname{J Low Temp Phys}

\DeclareMathOperator{\tr}{\mathop{\mathrm{Tr}}}

\DeclareMathOperator{\re}{\mathop{\mathrm{Re}}}

\newcommand{\Eq}[1]{Eq.~(\ref{#1})}
\newcommand{\Eqs}[1]{Eqs.~(\ref{#1})}
\begin{document}

\title{Nonequilibrium Electron Cooling by NIS Tunnel Junctions}

\author{A. S. Vasenko \and F. W. J. Hekking}

\institute{
A. S. Vasenko\and
F. W. J. Hekking\at
LPMMC, Universit\'{e} Joseph Fourier and CNRS, 25 Avenue des Martyrs, \\ BP 166, 38042 Grenoble, France\\
\email{Andrey.Vasenko@grenoble.cnrs.fr}\\
\email{Frank.Hekking@grenoble.cnrs.fr}}

\date{Received: date / Accepted: date}

\maketitle

\begin{abstract}
We discuss the theoretical framework to describe quasiparticle electric and heat currents in NIS tunnel junctions in the dirty limit. The approach is based on quasiclassical Keldysh-Usadel equations. We apply this theory to diffusive NIS$^\prime$S tunnel junctions. Here N and S are respectively a normal and a superconducting reservoirs, I is an insulator layer and S$^\prime$ is a nonequilibrium superconducting lead. We calculate the quasiparticle electric and heat currents in such structures and consider the effect of inelastic relaxation in the S$^\prime$ lead. We find that
in the absence of strong relaxation the electric current and the cooling power for voltages $eV < \Delta$ are suppressed.
The value of this suppression scales with the diffusive transparency parameter. We ascribe this suppression to the effect of backtunneling of nonequilibrium quasiparticles into the normal metal.
\end{abstract}

\section{Introduction}

It is well known that in NIS (Normal metal - Insulator - Superconductor) tunnel junctions
the flow of electric current carried by quasiparticles  is accompanied by a heat transfer from the normal metal into the
superconductor \cite{Nahum, Giazotto}. This happens due to the presence of the superconducting energy gap
$\Delta$, which induces selective tunneling of high-energy
quasiparticles out of the normal metal.
In the tunneling event only quasiparticles with energy $E > \Delta$ (compared to the Fermi level) can tunnel out of the
normal metal. They generate the single particle current and associated heat current. At lower energies $E < \Delta$
charge transfer occurs via mechanism of Andreev reflection \cite{Andreev, S-J}. Andreev current $I_A$ generates a Joule heating
$I_A V$ that is deposited in the normal metal electrode but this effect dominates over single particle cooling only at very low temperatures \cite{Sukumar1}. For the temperatures considered in this paper Andreev Joule heating is negligible.

It has been shown that for voltage-biased NIS tunnel
junctions heat current out of the normal metal (also referred as ``cooling power'') is positive when
$eV \lesssim \Delta$, i.e., it cools the normal metal \cite{LPA}.
For $eV \gtrsim \Delta$ the current through the junction increases strongly, resulting in Joule heating $IV$ and making the
heat current negative. The cooling power is maximal near $eV \approx \Delta$.

This effect enables the refrigeration of electrons in the normal metal. A microrefrigerator, based on an NIS tunnel
junction, has been first fabricated by Na\-hum {\it et al.} \cite{Nahum}. They have used a single NIS tunnel junction in order
to cool a small normal metal strip. Later Leivo {\it et al.} \cite{LPA} have noticed that the cooling
power of an NIS junction is an even function of an applied voltage, and have fabricated a refrigerator with
two NIS tunnel junctions arranged in a symmetric configuration (SINIS), which gives a
reduction of the electronic temperature from 300 mK to about 100 mK. This significant temperature reduction gives a perspective
to use NIS junctions for on-chip cooling of nano-sized systems like high-sensitivity detectors and quantum devices \cite{on-chip}.

To enhance the performance of the NIS refrigerator it is important to understand the role of possible
factors that may facilitate or decrease the cooling effect. One of such factors is the inelastic
relaxation of injected quasiparticles in the superconducting lead. In non-reservoir
geometries the quasiparticles injected into the superconducting lead generate a nonequilibrium distribution.
In a diffusive superconductor backscattering on impurities and subsequent backtunneling into normal metal may considerably reduce the net heat current out of the normal metal electrode.

The purpose of this work is to investigate the importance of the nonequilibrium quasiparticle distribution and
consider the effect of inelastic relaxation in the superconductor. Possible mechanisms of inelastic relaxation could
be usual processes, such as electron-electron or electron-phonon interactions, but also the presence of so
called ``quasiparticle traps'', i.e. additional normal metal electrodes connected to the superconductor,
which remove excited nonequilibrium quasiparticles from the superconducting lead \cite{traps, GV}. In this paper we do not specify the mechanisms
of inelastic relaxation and consider the relaxation time approximation approach.
Effect of the nonequilibrium quasiparticle injection in NIS junctions was also discussed in Ref.~\cite{Sukumar2}, where
the authors proposed a phenomenological model of quasiparticle diffusion in the superconducting lead.

The paper is organized as follows. In the next section, we formulate the theoretical model and basic equations.
In sections~\ref{Current} and \ref{Heat} we solve the kinetic equations and apply solutions for the calculation of the
electric and heat currents, respectively. In Sec.~\ref{Distrib} we calculate quasiparticle distribution functions in the
superconducting lead. Finally we summarize the results in Sec.~\ref{Conclusion}.


\section{Model and basic equations}\label{Model}

\begin{figure}[tb]
\epsfxsize=7cm\epsffile{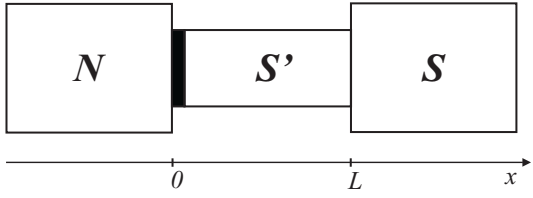} 
\caption{Geometry of the considered system.}
\label{model}
\end{figure}

The model of the N-I-S$^\prime$-S junction under consideration is
depicted in Fig.~\ref{model} and consists of a voltage-biased normal metal
reservoir (N), an insulator layer (I), a superconducting layer (S$^\prime$) of
thickness $L$ and a superconducting reservoir (S) along the $x$
direction. The S and S$^\prime$ leads are made from the identical superconducting material.
We assume the S$^\prime$-S interface to be fully transparent. We
will consider the diffusive limit, in which the elastic scattering
length $\ell $ is much smaller than the coherence length $\xi _0 =
\sqrt{\mathcal{D}/2\Delta}$, where $\mathcal{D}$ is the diffusion coefficient
(we assume $\hbar = k_B = 1$). The length $L$ of the S$^\prime$ lead is assumed
to be much larger than $\xi_0$. The problem of current flow through diffusive N-I-S$^\prime$-S
structures with short S$^\prime$ superconductor lead was solved in Ref.~\cite{Zaitsev}.

Under the conditions described above, the calculation of the electric and heat currents requires solution of the
one-dimensional Keldysh-Usadel equations \cite{LOnoneq} (see also review \cite{Belzig}) for the $4 \times 4$ matrix
Keldysh-Green function $\check{G}(x, E)$ in the S$^\prime$ lead,
\begin{align}
&\left[\check{\sigma}_z E + \check{\Delta}, \; \check{G}\right] = i \mathcal{D}
\partial \check{J}, \quad \check{J} = \check{G}\partial
\check{G}, \quad \check{G}^2 = \check{1},\label{GK}
\\ \label{G}
&\check{G} = \begin{pmatrix} \hat{g}^R & \hat{G}^K \\
0 & \hat{g}^A
\end{pmatrix}, \quad \hat{G}^K = \hat{g}^R \hat{f} - \hat{f}\hat{g}^A.
\end{align}
Here
\begin{equation}\nonumber
\check{\sigma}_z = \begin{pmatrix} \sigma_z & 0 \\ 0 & \sigma_z \end{pmatrix}, \quad
\check{\Delta} = \begin{pmatrix} \hat{\Delta} & 0 \\ 0 & \hat{\Delta} \end{pmatrix},
\end{equation}
$\hat{g}^{R,A}$ are the $2 \times 2$ Nambu matrix retarded and
advanced Green's functions, $\hat{f} = f_+ + \sigma_z f_-$ is the
matrix distribution function (we use ``check'' for $4 \times 4$ and
``hat'' for $2 \times 2$ matrices), $\sigma_{y,z}$ are the Pauli
matrices in the Nambu space, $\hat{\Delta} = e^{i \sigma_z \chi} i
\sigma_y \Delta$, $\Delta$ and $\chi$ are the modulus and the
phase of the pair potential, and $\partial \equiv \partial/\partial x$. In \Eqs{GK} we neglect the inelastic collision term
which will be taken into account later. Since we are interested in small voltages
$eV \lesssim \Delta$ (when the cooling of the normal metal occurs) we can neglect the effect of the suppression of
the superconducting gap due to the heating in the superconductor.

The boundary conditions for the function $\check{G}$ and the matrix
current $\check{J}$ at the left normal ($x=-0$) and the right superconducting ($x=+0$)
sides of the tunnel barrier are given by the relation \cite{KL}
\begin{equation}
\left( \sigma_N/g_N \right) \check{J}_{-0} = \check{J}_{+0} = (W/\xi_0)
\bigl[\check{G}_{-0}, \check{G}_{+0}\bigr],\label{KL}
\end{equation}
where $\sigma_N$ and $g_N$ are the normal conductivities of the N and S$^\prime$ leads per unit length,
respectively.
In \Eq{KL}, the transparency parameter $W$ is defined as
\begin{equation}\label{W}
W = R(\xi_0)/2R = (3\xi_0/4\ell)\Gamma \gg \Gamma,
\end{equation}
where $R(\xi_0) = \xi_0/g_N$ is the normal resistance of the S$^\prime$ lead per length $\xi_0$ and
$R$ is the junction resistance. It has been
shown in Refs.~\cite{Kupriyanov, Bezuglyi} that this quantity rather than the barrier
transparency $\Gamma$ plays the role of a transparency parameter
in the theory of diffusive tunnel junctions (see also the discussion in Ref.~\cite{paper_1}).
In this paper, we will consider the limit $W \ll 1$, which corresponds to the conventional tunneling concept.
At the right transparent S$^\prime$-S interface all functions and their first derivatives are to be continuous.
We neglect possible small resistance at this interface \cite{Nikolic} since it is much smaller than the resistance of the
S$^\prime$ lead in the normal state.

The electric and heat currents are related to the Keldysh
component of the matrix current $\check{J}$ respectively
as \cite{LOnoneq, BA, Heikkila1, Vinokur, Golubov},
\begin{align}\label{I}
I &= - \frac{g_N}{4e} \int_0^{\infty} \tr \sigma_z \hat{J}^K dE,
\\ \label{Q}
P &= IV + \frac{g_N}{4 e^2} \int_0^{\infty} E \tr \hat{J}^K dE,
\end{align}
and thus they can be expressed through the boundary value $\check{J}_{+0}$ in \Eq{KL},
\begin{align}\label{Ib}
I &= - \frac{1}{8 e R} \int_0^{\infty} \tr \sigma_z \bigl[\check{G}_{-0}, \check{G}_{+0}\bigr]^K dE,
\\ \label{Qb}
P &= IV + \frac{1}{8 e^2 R} \int_0^{\infty} E \tr \bigl[\check{G}_{-0}, \check{G}_{+0}\bigr]^K dE.
\end{align}

For further consideration it is convenient to write the Green
function in the following standard way,
\begin{equation}\label{g}
\hat{g} = \sigma_z u + i \sigma_y v.
\end{equation}
Here we neglect the phase of the anomalous Green function since it gives corrections
to the next order in diffusive barrier transparency parameter $W$.
The functions $u$ and $v$ determine the spectral characteristics of the system. In particular,
the quantity $N(E) = \left( u^R - u^A \right)/2$ is the density of states (DOS) normalized to its value $N_F$ in the
normal state. In what follows, we
will express the advanced Green functions through the retarded
ones, $(u,v)^{A} = - (u,v)^{R\ast}$, using the general relation
$\hat{g}^A = -\sigma_z \hat{g}^{R\dagger}\sigma_z$, and omit the
superscript $R$, considering retarded Green's functions only.

In this paper we will calculate the single particle current,
therefore we consider quasiparticle
energy $E > \Delta$ from now on. We neglect the proximity effect since proximity corrections to the spectral functions
are of the order of $W$. Therefore, we use the
BCS density of states in both S$^\prime$ and S layers. In the left voltage-biased
normal metal reservoir (N) we have,
\begin{equation}\label{g_N}
\hat{g}_N = \sigma_z, \quad f_{\pm N} = \frac{1}{2}\left[
\tanh\left(\frac{E+eV}{2T_N}\right) \pm
\tanh\left(\frac{E-eV}{2T_N}\right) \right],
\end{equation}
where $T_N$ is the temperature of the normal metal reservoir.
We assume the voltage $V$ to be directly applied to the tunnel barrier
and neglect a small electric field ($\sim eVW$) penetrating the S$^\prime$ superconducting lead.

In the right superconducting reservoir (S) Green's and distribution
functions are given by the relations
\begin{align}\label{g_S}
\hat{g}_S &= \sigma_z u_S + i \sigma_y v_S, \quad (u_S, v_S) = \frac{(E,\Delta)}{\sqrt{(E+i0)^2 - \Delta^2}},
\\ \label{f_S}
f_{+S}\bigl|_{x>L} &\equiv f_{eq} = \tanh\left( \frac{E}{2T_S}
\right), \quad f_{-S}\bigl|_{x>L} = 0,
\end{align}
where $T_S$ is the temperature of the S reservoir.

In the S$^\prime$ layer, Green's function is given by \Eq{g_S} and distribution
functions $f_{\pm S}(x,E)$ should be found from the kinetic equations,
which follow from the Keldysh component of \Eqs{GK} and for
$E > 0$ have a simple form $\partial^2 f_{\pm S} = 0$ within our approximations.
These equations have no bound solutions: both distribution functions $f_{\pm S}(x,E)$ grow
linearly with $x$ far from the junction. Such a growth is limited in practice by
inelastic collisions, which provide the spatial relaxation of $f_{\pm S}(x,E)$ to the
equilibrium values at $x \sim l_\pm \gg \xi_0$, where $l_\pm = \sqrt{\mathcal{D} \tau_\pm}$ are the inelastic
scattering lengths and $\tau_\pm$ are the inelastic scattering times. To simplify the problem,
instead of including complicated inelastic collision integrals, we add collision
terms in the relaxation time approximation to the kinetic equations,
\begin{align}\label{f_+}
l_+^2 \partial^2 f_{+S} &= (f_{+S} - f_{eq}) N(E),
\\ \label{f_-}
l_-^2 \partial^2 f_{-S} &=  f_{-S} / N(E),
\end{align}
where $N(E) = \re (u_S)$ is the BCS DOS.

We should supplement \Eqs{f_+}-\eqref{f_-} with proper boundary
conditions on both left and right interfaces. On the tunnel
barrier ($x = 0$) they follow from the Keldysh component of
\Eq{KL},
\begin{align}\label{KLf_+}
\partial_x f_{+S}\bigl|_{x=0} &= \frac{N(E)}{g_N R} \left( f_{+S0} - f_{+N} \right),
\\ \label{KLf_-}
\partial_x f_{-S}\bigl|_{x=0} &= \frac{1}{g_N R N(E)} \left( f_{-S0} - f_{-N} \right),
\end{align}
where $f_{\pm S0}(E)$ are the boundary values of $f_{\pm S}(x,E)$ at $x = 0$.
On the right transparent interface distribution functions become
equilibrium functions of the right S reservoir,
\begin{equation}\label{L}
f_{+S}\bigl|_{x=L} = f_{eq}, \quad f_{-S}\bigl|_{x=L} = 0.
\end{equation}
%


\section{Single particle current}\label{Current}

The equation for the electric current follows from \Eqs{Ib},
\eqref{KLf_-} and for single particle current reads,
\begin{equation}\label{I1}
I = \frac{1}{e R} \int_{\Delta}^{\infty} N(E) \left( f_{-N} - f_{-S0}
\right) dE.
\end{equation}
To obtain $f_{-S0}$ we should solve the boundary problem
\eqref{f_-},\eqref{KLf_-} and \eqref{L}.
Doing this we get the following result,
\begin{align}\label{f_-S0}
f_{-S0} &= f_{-N} \frac{\alpha_-}{\alpha_- +
\sqrt{N(E)}\coth(\beta_-/\sqrt{N(E)})},
\\ \label{alpha_minus}
\alpha_- &= 2W l_-/\xi_0 = R(l_-)/R, \quad \beta_- = L/l_-,
\end{align}
where $R(l_-) = l_-/g_N$ is the normal resistance of the superconducting lead per length $l_-$.
From \Eq{f_-S0} we see that the nonequilibrium correction to the current is of the order of $\alpha_- \gg W$,
which justifies our assumption about neglecting terms of the order of $W$ in the kinetic equation.

Substituting \Eq{f_-S0} into \Eq{I1} we finally obtain,
\begin{equation}\label{I_in}
I = \frac{1}{e R} \int_{\Delta}^{\infty} N(E) f_{-N}
\frac{\coth(\beta_-/\sqrt{N(E)})}{\alpha_-/\sqrt{N(E)} +
\coth(\beta_-/\sqrt{N(E)})} dE.
\end{equation}
For $L = 0$ or $l_- = 0$ the boundary value $f_{-S0}$ in \Eq{I1} is equal to zero and \Eq{I_in} reduces to the well-known
equation for single particle current between N and S reservoirs,
connected through an insulating layer,
\begin{equation}\label{I_eq}
I = \frac{1}{eR} \int_{-\infty}^{+\infty} N(E - eV) \left[
n_F(E-eV) - n_F(E) \right] dE,
\end{equation}
where $n_F(E) =[1 + \exp(E/T)]^{-1}$ is a Fermi function.

\begin{figure}[tb]
\epsfxsize=8.5cm\epsffile{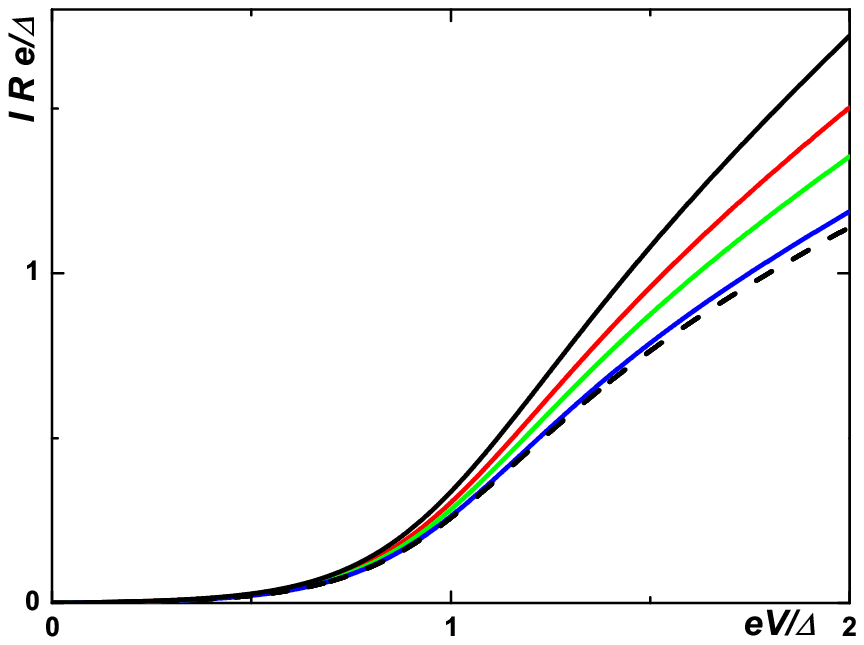} 
\caption{(Color online) IV characteristic of the NIS$^\prime$S junction.
$T_N$ = $T_S$ = 0.3 $T_C$, $W$ = 0.01. Solid black line is
calculated by use of \Eq{I_eq}. Other lines are calculated by use
of \Eq{I_in}, $L$ = 50$\xi_0$: $l_- = 10\xi_0$ (red line);
$l_- = 20 \xi_0$ (green line); $l_- = 50 \xi_0$
(blue line); $l_- = 100 \xi_0$ (dashed black
line).}
\label{curr_2}
\end{figure}

In the absence of inelastic relaxation in the S$^\prime$ lead ($\beta_- \ll 1$) we can approximate \Eq{I_in}
by
\begin{equation}\label{Iapp}
I = \frac{1}{e R} \int_{\Delta}^{\infty} f_{-N}
\frac{N^2(E)}{N(E) + \alpha_L} dE,
\end{equation}
where $\alpha_L = 2 W L/\xi_0 = R(L)/R$ and $R(L) = L/g_N$ is the resistance of the S$^\prime$ lead in the normal state.
At zero temperature at $eV \gg \Delta$ the current given by \Eq{Iapp} can be calculated to first order in the small
parameter $\Delta/eV \ll 1$ as follows,
\begin{equation}\label{assym}
I = \frac{1}{eR}\int_\Delta^{eV} \frac{N^2(E) \; dE}{N(E) +\alpha_L} = \frac{\Delta}{eR}\int_1^{eV/\Delta} \frac{x^2 \; dx}{x\sqrt{x^2 - 1} + \alpha_L (x^2 - 1)}
\approx \frac{V}{R_{tot}} + I_{exc},
\end{equation}
where $R_{tot} = R + R(L)$ is the net normal resistance of the junction and the S$^\prime$ lead and
$I_{exc}$ is an excess current given by the relation,
\begin{equation}
I_{exc} = \frac{\Delta}{eR} \left[ \frac{\alpha_L}{1 - \alpha_L^2} - \frac{2 \alpha_L^2}{\left( 1 - \alpha_L^2 \right)^{3/2}}
\arctan\left( \sqrt{\frac{1 - \alpha_L}{1 + \alpha_L}} \right)\right].
\end{equation}
Thus the IV characteristic \Eq{Iapp} exhibits a voltage-independent excess current at large voltage,
which is the manifestation of the nonequilibrium in the S$^\prime$ lead.
Here we should mention that the total excess current measured in the experiment is known to consist of the two
contributions: the one coming from the single particle current at large voltage just calculated above, and the other
coming from the two particle current (Andreev current). In this paper we do not calculate the latter contribution since
it is of the order of $W \ll \alpha_L$.

In Fig.~\ref{curr_2} we plot the IV characteristic of the NIS$^\prime$S
junction for different values of $l_-$ parameter.
I(V) given by \Eqs{I_in}, \eqref{I_eq} is an odd function of
voltage and we plot it only for positive voltages.
We fix $T_N = T_S = 0.3 T_C$, where $T_C$ is the critical temperature of the superconductor. For
Aluminum $T_N = T_S \approx 360$ mK.
We see that with the growth of the charge imbalance relaxation length $l_-$ the electric current decreases. When
$l_- > L$ the length of the S$^\prime$ lead $L$ plays the role of a characteristic relaxation length
and the current is almost independent of $l_-$.


\section{Cooling power}\label{Heat}

The equation for the cooling power follows from \Eqs{Qb},
\eqref{KLf_+} and reads
\begin{equation}\label{Q1}
P = - IV - \frac{1}{e^2 R} \int_{\Delta}^{\infty} E N(E) \left(
f_{+N} - f_{+S0} \right) dE,
\end{equation}
where $I$ is given by \Eq{I_in}. To obtain $f_{+S0}$ we should solve
boundary problem \eqref{f_+},\eqref{KLf_+} and \eqref{L}. Doing
this we get the following result,
\begin{align}\label{f_+S0}
f_{+S0} &= \frac{f_{+N} \alpha_+ \sqrt{N(E)} + f_{eq} \coth(\beta_+
\sqrt{N(E)})}{\alpha_+ \sqrt{N(E)} + \coth(\beta_+\sqrt{N(E)})},
\\ \label{alpha_plus}
\alpha_+ &= 2W l_+/\xi_0 = R(l_+)/R \gg W, \quad \beta_+ = L/l_+,
\end{align}
where $R(l_+) = l_+/g_N$ is the normal resistance of the superconducting lead per length $l_+$.

Substituting \Eq{f_+S0} into \Eq{Q1} we finally obtain,
\begin{equation}\label{Q_in}
P = - IV - \frac{1}{e^2 R} \int_{\Delta}^{\infty} E N(E) \left(
f_{+N} - f_{eq} \right) \frac{\coth(\beta_+\sqrt{N(E)})
dE}{\alpha_+\sqrt{N(E)} + \coth(\beta_+\sqrt{N(E)})}.
\end{equation}
For $L = 0$ or $l_\pm = 0$ the boundary value $f_{+S0}$ in \Eq{Q1} is equal to $f_{eq}$ and \Eq{Q_in} reduces to the well-known
equation for the heat current between N and S
reservoirs, connected through an insulating layer \cite{LPA},
\begin{equation}\label{Q_eq}
P = \frac{1}{e^2R} \int_{-\infty}^{+\infty} N(E) (E - eV) \left[
n_F(E-eV) - n_F(E) \right] dE.
\end{equation}

In the absence of inelastic relaxation in the S$^\prime$ lead ($\beta_\pm \ll 1$) we can approximate \Eq{Q_in}
by
\begin{equation}\label{Papp}
P = -IV - \frac{1}{e^2 R} \int_{\Delta}^{\infty} \left(
f_{+N} - f_{eq} \right) \frac{E N(E) dE}{1 + \alpha_L N(E)},
\end{equation}
where the current $I$ is given by \Eq{Iapp}.
This equation corresponds to the case when the length of the S$^\prime$ lead L is smaller than the inelastic relaxation length
and all relaxation occurs only in the S reservoir.

\begin{figure}[tb]
\epsfxsize=8.5cm\epsffile{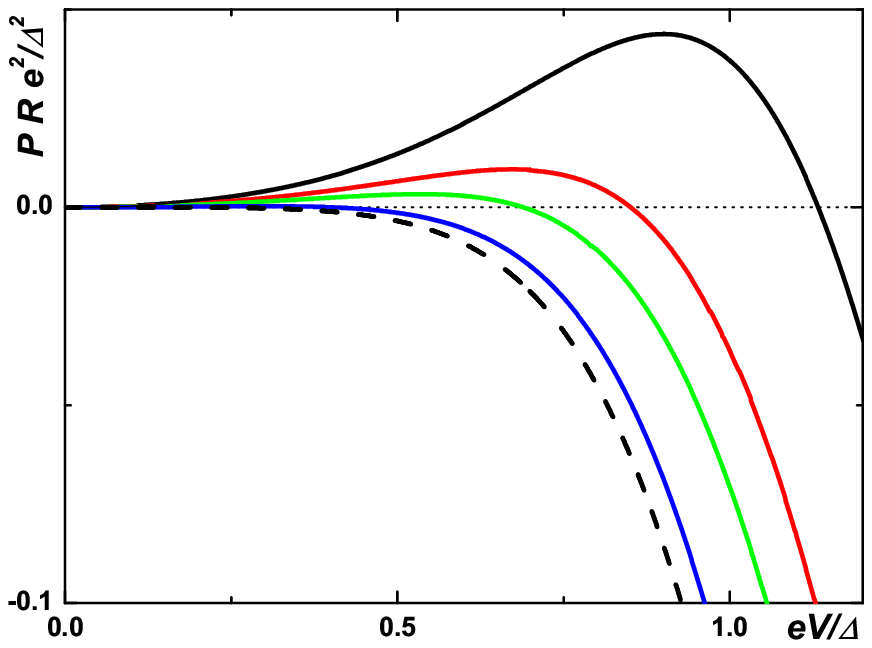} 
\caption{(Color online) P(V) dependence of the NIS$^\prime$S junction.
$T_N$ = $T_S$ = 0.3 $T_C$, $W$ = 0.01. Solid black line is
calculated by use of \Eq{Q_eq}. Other lines are calculated by use
of \Eq{Q_in}, $L$ = 50$\xi_0$: $l_- = l_+ = 10\xi_0$ (red line);
$l_- = l_+ = 20 \xi_0$ (green line); $l_- = l_+ = 50 \xi_0$
(blue line); $l_- = l_+ = 100 \xi_0$ (dashed black line).}
\label{heat_1}
\end{figure}
\begin{figure}[tb]
\epsfxsize=8.5cm\epsffile{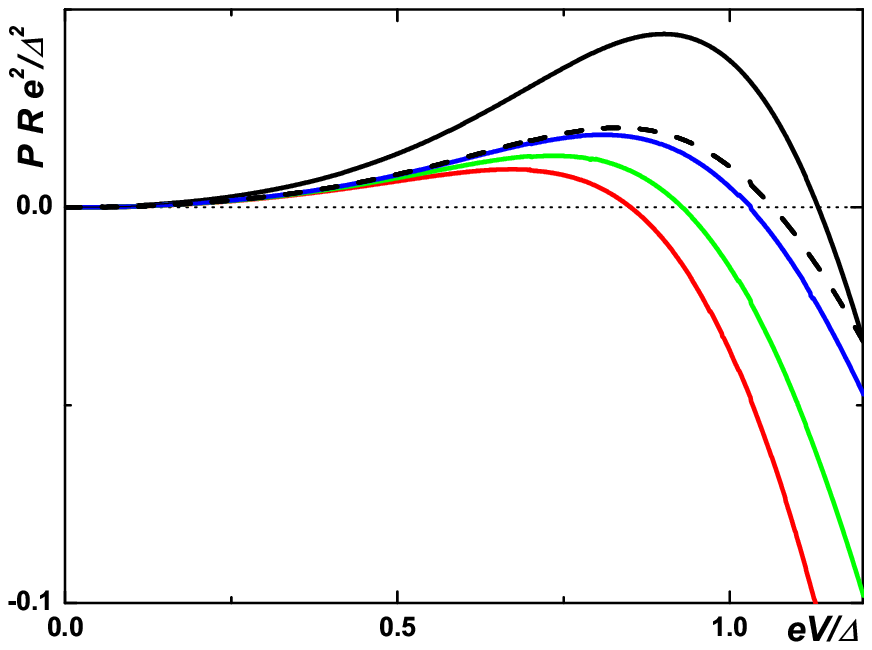} 
\caption{(Color online) P(V) dependence of the NIS$^\prime$S junction.
$T_N$ = $T_S$ = 0.3 $T_C$, $W$ = 0.01. Solid black line is
calculated by use of \Eq{Q_eq}. Other lines are calculated by use
of \Eq{Q_in}, $L$ = 50$\xi_0$, $l_+ = 10\xi_0$: $l_- = 10\xi_0$ (red line);
$l_- = 20 \xi_0$ (green line); $l_- = 50 \xi_0$
(blue line); $l_- = 100 \xi_0$ (dashed black line).}
\label{heat_2}
\end{figure}
\begin{figure}[tb]
\epsfxsize=8.5cm\epsffile{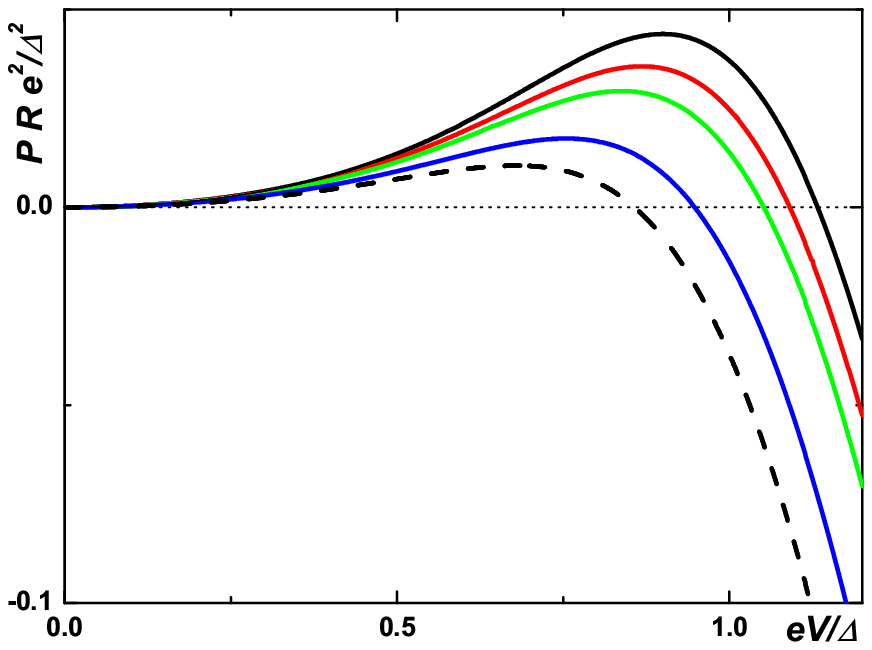} 
\caption{(Color online) P(V) dependence of the NIS$^\prime$S junction.
$T_N$ = $T_S$ = 0.3 $T_C$, $W$ = 0.001. Solid black line is
calculated by use of \Eq{Q_eq}. Other lines are calculated by use
of \Eq{Q_in}, $L$ = 50$\xi_0$: $l_- = l_+ = 10\xi_0$ (red line);
$l_- = l_+ = 20 \xi_0$ (green line); $l_- = l_+ = 50 \xi_0$
(blue line); $l_- = l_+ = 100 \xi_0$ (dashed black line).}
\label{heat_3}
\end{figure}

In Figs.~\ref{heat_1},\ref{heat_2},\ref{heat_3} we plot the P(V) dependence  of the NIS$^\prime$S
junction for different values of $l_\pm$ parameters.
P(V) given by \Eqs{Q_in}, \eqref{Q_eq} is an even function of
voltage and we plot it only for positive voltages.
From Fig.~\ref{heat_1} it can be seen that with the growth of the relaxation lengths $l_\pm$ the cooling power decreases.
Also the maximal cooling power shifts to the region of smaller voltages than in the equilibrium case.
We ascribe this suppression to the effect of backscattering on impurities and tunneling of nonequilibrium quasiparticles back to the
normal metal reservoir. We can see that for large values of $l_\pm$ the cooling power is negative for all voltages.

In Fig.~\ref{heat_2} we plot P(V) for different ratio $l_+/l_-$. For a fixed $l_+$ length we vary the charge imbalance
relaxation length $l_-$. It can be seen that the cooling power increases with the growth of $l_-$. This happens because of the decrease
of the term $IV$ in \Eq{Q_in} due to the suppression of the electric current (see Sec.~\ref{Current}).

Finally we want to stress here the role of the transparency parameter $W$. In Fig.~\ref{heat_3} we plot the P(V) dependence
for a different value of tunneling parameter $W = 10^{-3}$, which is one order of magnitude smaller than $W$ in
Figs.~\ref{heat_1},\ref{heat_2}. Here we again see the decrease of the cooling power with the growth of $l_\pm$, but
the effect is smaller than in Fig.~\ref{heat_1}. This is obvious since the amplitudes of the nonequilibrium distribution functions
$f_{\pm S0}$ scale with the $W$ parameter. For very strong tunnel barriers the nonequilibrium
effect is therefore negligible.

In order to estimate the characteristic parameters of the junctions for the transparency parameters $W = 10^{-2}$
and $W = 10^{-3}$ used in our numerical calculations, we will assume the junction area to be $200 \times 200$
nm and the thickness of the leads as well as the mean free path to be 50 nm. For Al leads, this results in the sheet resistance
$R_{\Box} \approx 0.3$ $\Omega$ and
$R(\xi_0)\approx 0.45$ $\Omega$ at $\xi_0 \approx 300$ nm. Then,
according to \Eq{W}, the tunneling probability and the junction
resistance approach the values $\Gamma
\approx 2 \times 10^{-3}$, $R \approx 22.5$ $\Omega$ for $W = 10^{-2}$, and  $\Gamma
\approx 2 \times 10^{-4}$, $R \approx 225$ $\Omega$ for $W = 10^{-3}$, respectively.


\section{Nonequilibrium quasiparticle distribution in S$^\prime$}\label{Distrib}

\begin{figure}[tb]
\epsfxsize=11cm\epsffile{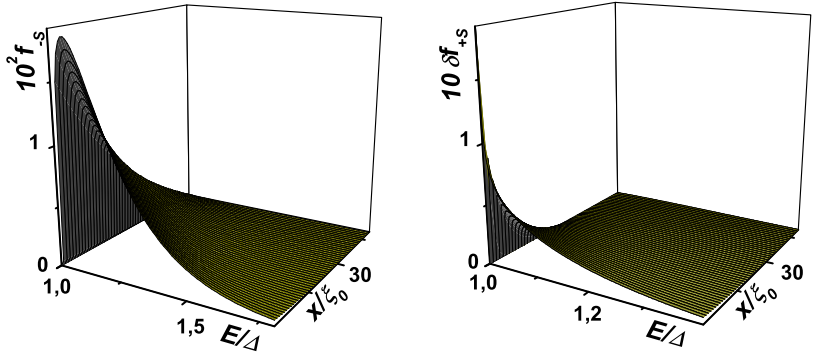} \vspace{50mm}
\caption{(Color online) Distribution functions $f_{-S}(E,x)$ (left plot) and
$\delta f_{+S}(E,x) = f_{eq} - f_{+S}$ (right plot). Here $T_N$ = $T_S$ = 0.3 $T_C$, $eV = 0.8 \Delta$,
$L = 50 \xi_0$, $l_- = l_+ = 10 \xi_0$, $W$ = 0.01.}
\label{DMP}
\end{figure}

Solving \Eqs{f_-},\eqref{KLf_-} and \eqref{L} we can obtain the function $f_{-S}(x,E)$ in the S$^\prime$ lead. It reads
\begin{equation}\label{f_-(x)}
f_{-S} = \frac{f_{-N}\alpha_-}{\alpha_- + \sqrt{N(E)}\coth\left(\beta_-/\sqrt{N(E)}\right)} \frac{\sinh\left[ \left( \beta_- - x/l_- \right)/\sqrt{N(E)} \right]}{\sinh\left( \beta_-/\sqrt{N(E)} \right)}.
\end{equation}
This equation describes the relaxation of the imbalance between the electrons and holes in the S$^\prime$ lead (branch mixing). From \Eq{f_-(x)} we see the exponential
decay of the imbalance function $f_-$ at the distance $x \sim l_-\sqrt{N(E)}$ from the tunnel barrier. At $E \rightarrow \Delta$
the decay length diverges.

Similarly from \Eqs{f_+},\eqref{KLf_+} and \eqref{L} we obtain the function $f_{+S}(x,E)$ in the S$^\prime$ lead,
\begin{equation}
f_{+S} = f_{eq} - \frac{\left( f_{eq} - f_{+N} \right)\alpha_+ \sqrt{N(E)}}{\alpha_+ \sqrt{N(E)} + \coth\left( \beta_+ \sqrt{N(E)} \right)} \frac{\sinh\left[ \left( \beta_+ - x/l_+ \right)\sqrt{N(E)} \right]}{\sinh\left(\beta_+\sqrt{N(E)}\right)}.
\end{equation}
From this equation we see the exponential decay of the $\delta f_{+S}$ function at the distance $x \sim l_+/\sqrt{N(E)}$ from the tunnel barrier, where $\delta f_{+S} = f_{eq} - f_{+S}$.

We note that the charge imbalance function $f_-$ can be measured in the experiment \cite{distrib}.
In the adopted geometry, Fig.~\ref{model}, one can probe the $f_{-S}$ function at
different locations in S$^\prime$ by small normal metal electrodes N$_i^\prime$, connected to S$^\prime$ through
the insulating barrier at different points $x_i$. Importantly, the area
of these probing N$_i^\prime$IS$^\prime$ junctions should be much smaller than the area of the initial NIS$^\prime$ barrier. Driving the
current through the probing junction one can measure the corresponding IV curve and calculate the
function $f_{-S}$ in the S$^\prime$ lead.

In Fig.~\ref{DMP} we plot both $f_{-S}(E,x)$ and $\delta f_{+S}(E,x)$ functions. We can see that nonequilibrium distributions
are nonzero only at energies rather close to $\Delta$. For large energies $E \gg \Delta$ both functions are approaching zero.

In this paper we used simple relaxation approximation approach. At zero temperature it gives qualitatively correct
estimation of nonequilibrium properties of considered systems (see, for example, Ref.~\cite{Golubov2}).
We note that this approximation leads to an incorrect expression of the charge imbalance relaxation length near the critical
temperature \cite{Volkov}. At intermediate temperatures $0 < T < T_C$ this approximation gives a result correct with an accuracy of a numerical factor.
Since in this paper we consider the temperatures $T_N = T_S$ equal to $0.3 T_C$ which is much smaller than $T_C$ our description is
qualitatively correct. However, we should stress that for higher temperatures one should add inelastic collision integrals
in the kinetic equations \cite{Kaplan, Heikkila2, Kopnin}.


\section{Conclusion}\label{Conclusion}

We investigated the electric and heat currents in an NIS$^\prime$S junction in the diffusive limit. We have developed a model
which describes the nonequilibrium quasiparticle injection and relaxation in the superconducting lead. This model
will be used as a tool to fit experimental data in various types of NIS tunnel junctions in non-reservoir geometries.
These fits will be presented elsewhere.

We showed that in the case when relaxation lengths in the superconductor are rather long compared to the coherence length, the electric current and
the cooling power for $eV$ below $\Delta$ are suppressed. We ascribe this suppression to the backtunneling of nonequilibrium quasiparticles into the
normal metal. The value of this suppression scales with the diffusive transparency parameter $W$.

Finally, we calculated the nonequilibrium distribution functions in the superconducting lead.

\begin{acknowledgements}
The authors thank D.~V.~Averin, E.~V.~Bezuglyi, H.~Courtois, D.~S.~Golubev, A.~A.~Golubov, T.~T.~Heikkil\"{a}, M.~Houzet, S.~Rajauria and A.~F.~Volkov for useful discussions. This work was supported
by Na\-no\-SciERA ``Na\-no\-fridge'' EU project.
\end{acknowledgements}

\end{document}